\documentclass[prb,showpacs,twocolumn,superscriptaddress]{revtex4}
\usepackage{hyperref,graphicx,dcolumn}
\usepackage{amsfonts,amsmath,amssymb,bm,bbm,mathtools,siunitx}
\usepackage{color}
\usepackage{pdfsync}
\usepackage{blkarray}
\usepackage{multirow}
\usepackage{braket}
\usepackage{lipsum}
\usepackage{verbatim}

\newcommand{\myFig}[7]{ %
\begin{figure}[htb] 
\begin{center} 
\includegraphics[width=#1\columnwidth,height=#2\columnwidth,clip=true,keepaspectratio=#3,angle=#4]{#5}
\caption{#6} \vspace{-0.5cm} \label{#7} 
\end{center} \end{figure}}

\newcommand{\abs}[1]{\left\vert #1 \right\vert}
\newcommand{\Norm}[1]{\| #1 \|}

\begin{document}
\title{Ultrafast demagnetizing fields from first principles}
\author{Jacopo Simoni}\email[Contact email address: ]{simonij@tcd.ie}
\affiliation{School of Physics, AMBER and CRANN Institute, Trinity College, Dublin 2, Ireland} 
\author{Maria Stamenova} 
\affiliation{School of Physics, AMBER and CRANN Institute, Trinity College, Dublin 2, Ireland} 
\author{Stefano Sanvito} 
\affiliation{School of Physics, AMBER and CRANN Institute, Trinity College, Dublin 2, Ireland}

\begin{abstract}
We examine the ultrafast demagnetization process of iron-based materials, namely Fe$_6$ clusters and bulk bcc Fe, with 
time-dependent spin-density functional theory (TDSDFT). The magnetization continuity equation is reformulated and the 
torque due to the spin-current divergence is written in terms of an effective time-dependent {\it kinetic} magnetic field, an 
object already introduced in literature before. Its time 
evolution, as extracted from the TDSDFT simulations, is identified as one of the main sources of the local out-of-equilibrium 
spin dynamics and plays a major role in the demagnetization process. Such demagnetization is particularly strong in ``hot spots'' 
where the kinetic torque is maximized. Finally, we find the rate of demagnetization in Fe$_6$ to be strongly dependent on the 
direction of polarization of the exciting electric field and this can be linked to the out of equilibrium distribution of the
kinetic field in two comparative cases.    
\end{abstract}


\maketitle

\section{Introduction}

The search for practical solutions for increasing the speed of manipulation of magnetic bits is essential for the progress of 
modern information and communication technology. It has been shown that there is an upper limit to the speed of the magnetization 
switching process when this is driven by a magnetic field~\cite{Mag1, Mag2}. An increase in power absorption beyond this limit and 
for higher magnetic field amplitudes push a spin system out of equilibrium into a chaotic behaviour, and the switching speed decreases. 
For this reason the discovery made by Beaurepaire {\it et al.} [\onlinecite{Beau96}] in 1996 that a ferromagnetic Ni film could be 
demagnetized by a $60$ femtosecond optical laser pulse attracted a lot of interest and was the seed to a new field, now called 
femto-magnetism. 

In a standard pump-probe experiment the system is initially excited by an optical pulse (pump) and then the magnetization dynamics 
is monitored by analysing a second signal (probe)~\cite{Rasing10,Kimel}. Depending on the minimal delay between the pump and
the probe, one can analyse the demagnetization process at different timescales and thus observe the dissipation mechanisms active 
at that particular time. The interpretation of the results is, however, a complicate matter. In general for demagnetization processes observed 
on a timescale ranging from nanoseconds to $100$ picoseconds one considers an empirical three temperature model~\cite{3Temp},
where electrons, spins and phonons define three energy baths, all interacting with each other. In contrast, ultrafast spin dynamics, 
taking place within a few hundreds femtoseconds, is yet not described in terms of a single unified scheme and various models for the 
demagnetization process have been advanced. These include fully relativistic direct transfer of angular momentum from the light to
the spins~\cite{ultrarel, ultrarel2}, dynamical exchange splitting~\cite{dynEXCH}, electron-magnon spin-flip scattering~\cite{elmag}, 
electron-electron spin-flip scattering~\cite{elel} and laser-generated superdiffusive spin currents~\cite{superdiff}.

Given the complexity of the problem {\it ab initio} methods, resolved in the time domain, provide a valuable tool to probe the 
microscopic aspects of the ultrafast spin dynamics of real magnetic materials by means of time-dependent simulations. In this work 
we apply time-dependent spin density-functional-theory (TD-SDFT)~\cite{TDDFT, TDDFT2} in its semi-relativistic, non-collinear, 
spin-polarized version to analyse the ultrafast laser-induced demagnetization of two ferromagnetic transition metal systems: a 
Fe$_6$ cluster (see Fig.~\ref{fig:01}) and bulk bcc F\lowercase{e}. Recently, within a similar theoretical description, it has been demonstrated 
that the spin-orbit (SO) interaction plays a central role in the demagnetization process~\cite{Krieger15, Tows, Zhang}. Furthermore, 
it was showed by us~\cite{us} that the laser-induced spin dynamics can be understood as the result of the interplay between the SO 
coupling potential and an effective magnetic field. The so-called {\it kinetic magnetic field} \cite{antr1, antr2}, 
$\mathbf{B}_\mathrm{kin}(\mathbf{r},t)$, originates from the presence of non-uniform spin currents in the system. In this work we focus 
on the anatomy of $\mathbf{B}_\mathrm{kin}(\mathbf{r},t)$ and we analyze in detail its role in the highly non-equilibrium process of ultrafast 
demagnetization.   

The first formulation of the spin dynamics problem in transition metal systems was given in Refs.~[\onlinecite{antr1, antr2}] by 
Antropov and Katsnelson, who laid down the foundation of DFT-based spin dynamics, by deriving a set of equations of motion 
for the local magnetization vector. In those seminal works the magnetization dynamics was analyzed at the level of the adiabatic 
local spin-density approximation (ALSDA), but actual applications to real out-of-equilibrium systems were not described. Our 
purpose is to clarify and quantify, through TDSDFT simulations at the level of the non-collinear ALSDA, the role played by 
$\mathbf{B}_\mathrm{kin}(\mathbf{r}, t)$ in the laser-induced ultrafast spin dynamics of transition metal ferromagnets. 

The paper is divided into four main sections. In Section \ref{theory} we define the various fields that couple to the spins by 
isolating in the continuity equation only the terms that play a major role in the dynamical process. In Section \ref{res} we present 
the results of the calculations for F\lowercase{e}$_6$ clusters and show that ``hot spots'' for demagnetization are associated with 
larger misalignment of the kinetic magnetic field and the local spin density. This becomes more clear through evaluation of material 
derivatives. A demonstration of the effect of the polarization of the electric field on the rate of demagnetization of Fe$_6$ is 
discussed in Section \ref{Edir}. In Section \ref{bccFe} we show that previous observations for F\lowercase{e}$_6$ are valid for bulk bcc 
F\lowercase{e} as well. Finally we conclude. The paper is supplemented with an Appendix where we present a detailed derivation 
of the spin continuity equation (\ref{app}).  

\section{Theory} \label{theory}
We consider the TDSDFT problem within the ALSDA for a spin-polarized system excited by an electric field pulse. 
If one neglects second-order contributions arising from the solution of the coupled Maxwell-Schr\"odinger system 
of equations, the dynamics will be governed by the usual set of time-dependent Kohn-Sham (KS) equations
\begin{equation}\label{TDKS}
 i\hbar\frac{d}{dt}\psi_{j}^\mathrm{KS}(\mathbf{r}, t) = H_\mathrm{KS}(\mathbf{r}, t)\psi_{j}^\mathrm{KS}(\mathbf{r}, t)\:.
\end{equation}
In Eq.~(\ref{TDKS}) $\psi_{j}^\mathrm{KS}(\mathbf{r}, t)$ are the KS orbitals and the KS Hamiltonian, $H_\mathrm{KS}(\mathbf{r},t)$, 
can be expressed in the velocity gauge formulation and the minimal coupling substitution as,
\begin{align}\label{HKS}
H_\mathrm{KS}(\mathbf{r}, t) = & \frac{1}{2m}\Big( -i\hbar\nabla - \frac{q}{c}\mathbf{A}_\mathrm{ext}(t)\Big)^{2} - \nonumber \\
& -\mu_\mathrm{B}\hat{\bm{\sigma}}\cdot\mathbf{B}_\mathrm{s}[n, \mathbf{m}](\mathbf{r}, t) + v_\mathrm{s}[n](\mathbf{r}, t)\:, 
\end{align}
where
\begin{eqnarray}  
 v_\mathrm{s}[n](\mathbf{r}, t) & \!\!=\!\!& \int d^{3}\mathbf{r}^\prime \frac{n(\mathbf{r}^\prime)}{\abs{\mathbf{r} - \mathbf{r}^\prime}} 
 + v_\mathrm{xc}^\mathrm{ALSDA}[n](\mathbf{r}, t) + \nonumber \\ 
 & & + \sum_{I} V^{I}_\mathrm{PP}(\abs{\mathbf{r} - \mathbf{R}_{I}})
\end{eqnarray}
and
\begin{eqnarray}   
\mathbf{B}_\mathrm{s}[n, \mathbf{m}](\mathbf{r}, t) & \!\!=\!\!& \mathbf{B}_\mathrm{xc}^\mathrm{ALSDA}[n, \mathbf{m}](\mathbf{r}, t) + \mathbf{B}_\mathrm{ext}(\mathbf{r}, t)\:.\label{eq:Bs}
\end{eqnarray}

Here $v_\mathrm{s}(\mathbf{r}, t)$ represents the usual non-interacting KS potential and the full non-interacting magnetic field, 
$\mathbf{B}_\mathrm{s}(\mathbf{r}, t)$, consists of the external one, $\mathbf{B}_\mathrm{ext}(\mathbf{r}, t)$, and the 
exchange-correlation (XC) magnetic field, $\mathbf{B}_\mathrm{xc}^\mathrm{ALSDA}(\mathbf{r}, t)$. In the equations 
above $m$ is the electron mass, $q$ the electron charge, $c$ the speed of light, $\mathbf{A}_\mathrm{ext}(t)$ the vector 
potential associated to the external magnetic field, $\hat{\bm{\sigma}}$ the spin operator, $\mu_\mathrm{B}$ the Bohr 
magneton, $n$ the electron density and $\mathbf{m}$ the magnetization density. Then, $v_\mathrm{s}(\mathbf{r}, t)$ is 
decomposed into a Hartree contribution, an XC correlation one, $v_\mathrm{xc}^\mathrm{ALSDA}[n](\mathbf{r}, t)$, and 
into an ionic pseudo-potential $V^{I}_\mathrm{PP}(\abs{\mathbf{r} - \mathbf{R}_{I}})$. For a fully relativistic, 
norm-conserving pseudopotential the SO coupling enters into the KS equations in the form~\cite{SOC1}
\begin{equation}\label{eq:VSO}
\begin{split}
V^{I}_\mathrm{PP}(\abs{\mathbf{r} - \mathbf{R}_{I}}) & = \sum_{l} \Big( \bar{V}_{l}^{I}(\mathbf{r}) + \frac{1}{4}V_{l}^{I,\mathrm{SO}}(\mathbf{r}) +  \\
 & + \sum_{m=-l}^{l} V_{l}^{I,\mathrm{SO}}(\mathbf{r}) \hat{\mathbf{L}}_{I}\cdot \hat{\mathbf{S}} \ket{I, l, m} \bra{I, l, m} \Big).
\end{split}
\end{equation}
In Eq.~(\ref{eq:VSO}) the orbital momentum operator associated to the $I$-th atomic center is $\hat{\mathbf{L}}_{I}$, while 
the vectors $\{\ket{I, l, m}\}$ are the associated set of spherical harmonics centered on that given atomic position. In 
Eq.~(\ref{eq:VSO}) $V_{l}^{I,\mathrm{SO}}(\mathbf{r})$ defines a generalized space-dependent SO coupling parameter 
providing a measure of the SO interaction strength close to the atomic site, while $\bar{V}_{l}^{I}(\mathbf{r})$ includes all 
the ionic relativistic corrections like the Darwin and the mass correction term. Within the ALSDA $v_\mathrm{xc}(\mathbf{r},t)$ 
and $\mathbf{B}_\mathrm{xc}(\mathbf{r},t)$ are local functions in time of the electron density and magnetization, which in 
turn are written in terms of the time-dependent KS orbitals
\begin{eqnarray}
 n(\mathbf{r},t) &=& \sum_{j\in\mathrm{occ.}}\sum_{\sigma} \psi_{j\sigma}^\mathrm{KS}(\mathbf{r},t)^{*} \psi_{j\sigma}^\mathrm{KS}(\mathbf{r},t)\:, \\
 \mathbf{m}(\mathbf{r},t) &=& \sum_{j\in\mathrm{occ.}}\sum_{\alpha,\beta} \psi_{j\alpha}^\mathrm{KS}(\mathbf{r},t)^{*} \bm{\sigma}_{\alpha,\beta} \psi_{j\beta}^\mathrm{KS}(\mathbf{r},t)\:.
\end{eqnarray}
Starting from the set of time-dependent KS equations in (\ref{TDKS}) it is possible to derive an equation of motion for the 
magnetization, or a spin-continuity equation, in terms of the non-interacting KS observables. This reads
\begin{align}\label{Eq:SpinCon}
 \frac{d}{dt}\mathbf{m}(\mathbf{r},t) & = -\nabla \cdot \mathbf{J}_\mathrm{KS}(\mathbf{r}, t) 
 + \mu_\mathrm{B} \mathbf{m}(\mathbf{r}, t)\times \mathbf{B}_\mathrm{s}(\mathbf{r}, t) + \nonumber \\
& + \mathbf{T}_\mathrm{SO}(\mathbf{r}, t)\:,
\end{align}
where $\mathbf{J}_{\mathrm{KS}}(\mathbf{r},t)$ represents the non-interacting KS spin-current rank-$2$ tensor
\begin{equation}
 \mathbf{J}_\mathrm{KS}(\mathbf{r}, t) = 
 \frac{\hbar}{2mi} \sum_{j\in\mathrm{occ.}}\big(\psi_j^{\mathrm{KS}\dagger}\hat{\boldsymbol{\sigma}}\nabla\psi_j^{\mathrm{KS}} - \mathrm{h.c.}\big)\:,
\end{equation}
and the SO torque contribution reads
\begin{align}\label{Eq:SOtorque}
 \mathbf{T}_\mathrm{SO}(\mathbf{r}, t) & = \sum_{I}\sum_{l,m_{1},m_{2}}\sum_{j,\alpha,\beta}^\mathrm{occupied} V_{l}^\mathrm{SO}(\abs{\mathbf{r} - \mathbf{R}_{I}}) \cdot \nonumber \\
 & \cdot\braket{\psi^\mathrm{KS}_{j\alpha}|l,m_{1},I} \bra{l,m_{1},I}\mathbf{L}_{I}\ket{l,m_{2},I}\times\bm{\sigma}_{\alpha\beta} \cdot \nonumber \\
 & \cdot\braket{l,m_{2},I|\psi^\mathrm{KS}_{j\beta}}.
\end{align}
The KS magnetic field $\mathbf{B}_\mathrm{s}(\mathbf{r}, t)$ is taken as in Eq.~(\ref{eq:Bs}), which in absence of an 
external magnetic field reduces to $\mathbf{B}_\mathrm{xc}(\mathbf{r}, t)$. In DFT there are a set of zero-force theorems 
stating that the interaction between the particles cannot generate a net force \cite{levy}. In the case of the exchange-correlation 
magnetic field we have the exact condition $\int d^{3}r\:\mathbf{m}(\mathbf{r}, t)\times \mathbf{B}_\mathrm{xc}(\mathbf{r}, t)=0$, 
which is satisfied by the ALDA. Combining this equality with the assumption that the currents at the system boundary are 
negligible allows us to conclude that the only source of global spin loss is the SO coupling torque, $\mathbf{T}_\mathrm{SO}$, 
and that the spin lost during the temporal evolution is transferred to the orbital momentum of the system, which in turn is partially 
damped into the lattice (we consider frozen ions). Hence we have the relation,
\begin{equation}\label{eq:Tso}
 \frac{d}{dt}\int_{\Omega} d^3r\: \mathbf{m}(\mathbf{r}, t) = \int_{\Omega}d^3r\: \mathbf{T}_\mathrm{SO}(\mathbf{r}, t)\:,
\end{equation}
where the integration extends over the entire volume $\Omega$.

Within the ALDA, the exchange-correlation functional satisfies also a local variant of the zero-torque theorem~\cite{SpinDFT}, 
which is not a property of the exact DFT functional~\cite{Sharma07, Eich13, Sharma07b}. According to this condition 
$\mathbf{m}(\mathbf{r}, t)\times \mathbf{B}_\mathrm{xc}(\mathbf{r}, t) = 0$ 
and therefore the exchange-correlation magnetic field cannot contribute, even locally, to the magnetization dynamics. This leads 
us to conclude that the local magnetization dynamics is solely the result of the interplay between the spin-polarized currents and 
the SO torque (in reality $\mathbf{B}_{\mathrm{xc}}$ can still contribute indirectly to the spin dynamics through a dynamical 
modification of the gap between up and down spin polarized bands, which in turn determines an enhancement of the spin 
dissipation via the spin orbit coupling channel). In order to elucidate this view further we make use of the hydrodynamical 
formalism applied to spin systems, which has been already introduced in References~[\onlinecite{hydro2, hydro3}]. This 
approach needs to be slightly modified in view of the fact that we are considering an effective Kohn-Sham system and not 
a set of independent spin particles. In fact, as it was already pointed out in Refs.~[\onlinecite{antr1, antr2}], 
Eq.~(\ref{Eq:SpinCon}) can be written in a different form (the details of the derivation are shown in the appendix \ref{app})
\begin{align}\label{eq:spincon}
 \frac{D}{Dt}\mathbf{m}(\mathbf{r}, t) & + \sum_{j\in\mathrm{occ.}}\nabla\cdot\mathbf{v}_j(\mathbf{r},t) \mathbf{m}_j(\mathbf{r}, t) = -\nabla\cdot\mathcal{D}(\mathbf{r},t) + \nonumber \\
 & + \mu_\mathrm{B} \mathbf{m}(\mathbf{r}, t)\times\mathbf{B}_\mathrm{eff}(\mathbf{r}, t) + \mathbf{T}_\mathrm{SO}(\mathbf{r}, t)\:,
\end{align}
where a couple of new terms appear. In the equation $\frac{D}{Dt}=\frac{d}{dt}+\mathbf{v}\cdot\nabla$ is a material 
derivative, $\mathbf{v}_j(\mathbf{r}, t)$ represents a single Kohn-Sham state velocity field (see appendix \ref{app}), 
and $\mathbf{m}_j(\mathbf{r},t)=\psi_j^{\mathrm{KS}\dagger}\hat{\boldsymbol{\sigma}}\psi_j^{\mathrm{KS}}$. 
On the right hand-side of Eq.~(\ref{eq:spincon}) in addition to the spin-orbit coupling torque, $\mathbf{T}_{\mathrm{SO}}(\mathbf{r},t)$, 
we have a new term, $-\nabla\cdot\mathcal{D}(\mathbf{r},t)$, that describes the spin dissipation in the system due to the 
internal motion of the spin currents. It can be interpreted as an effective spin-current divergence object involving only transitions 
among different Kohn-Sham states [inter-band transitions, see Eq.~(\ref{Eq:spindiss})].
Finally the effective field $\mathbf{B}_{\mathrm{eff}}$ is given by the sum of two terms, 
$\mathbf{B}_{\mathrm{eff}}=\mathbf{B}_{\mathrm{xc}}+\mathbf{B}_{\mathrm{kin}}$, with $\mathbf{B}_{\mathrm{xc}}$
exchange-correlation field and $\mathbf{B}_{\mathrm{kin}}$ defined as [see Eq.~(\ref{Eq:Beff})]
\begin{equation}\label{Eq:Bkin}
 \mathbf{B}_\mathrm{kin}(\mathbf{r},t) = \frac{1}{\bar{\mathcal{F}}e}\bigg[ \frac{\nabla n\cdot\nabla\mathbf{s}}{n} + \nabla^2\mathbf{s}\bigg]\:,
\end{equation}
with spin vector field $\mathbf{s}(\mathbf{r}, t) = \frac{\mathbf{m}(\mathbf{r},t)}{n(\mathbf{r},t)}$.

Such $\mathbf{B}_\mathrm{kin}(\mathbf{r}, t)$ field has only an instrumental r\^ole in the equations of motion for 
the spin density, a very similar expression was already introduced in some previous work. In Ref.~[\onlinecite{antr1}] 
it was expressed in the form $\partial_k\frac{1}{n}(\mathbf{m}\times\partial_k\mathbf{m})$, while in Ref.~[\onlinecite{antr2}] 
it appears as $\frac{\nabla n\nabla\mathbf{m}}{n}$. The interpretation of $\mathbf{B}_{\mathrm{kin}}$
may look quite obscure at a first sight, however, in Ref.~[\onlinecite{Taka55, Taka57}] it was identified as a possible source 
of spin wave excitations in the form of a spin-spin interaction potential. 

In order to clarify this point, let us consider the Heisenberg interaction between two spins centered on atoms placed at a distance $d=|\mathbf{d}|$. 
We can assume naively, but reasonably, that the spin-spin interaction among the two spin distributions, computed at an arbitrary point $\mathbf{r}$ in space could be expressed in the following form
\begin{equation}
 H_\mathrm{eff}(\mathbf{r}) \, \simeq \, \mathbf{s}(\mathbf{r} -\mathbf{d}/2)\cdot \mathbf{s}(\mathbf{r} +\mathbf{d}/2)\:,
\end{equation}
where it is more convenient for us to employ a spin field, $\mathbf{s}(\mathbf{r})$, which describes the spin distribution in 
space, instead of an atom localized spin vector. Hence, $H_{\mathrm{eff}}$ defines an effective single-particle Hamiltonian. 
By averaging over the number of electrons in the entire space we obtain
\begin{equation}
 \mathbf{S}_{1}\cdot \mathbf{S}_{2} \simeq \int_{\Omega} d^{3}r \: n(\mathbf{r}) \mathbf{s}(\mathbf{r} -\mathbf{d}/2)\cdot\mathbf{s}(\mathbf{r} +\mathbf{d}/2)\:.
\end{equation}
Then, by expanding the spin density in Taylor series up to second order in the distance $d$ and by neglecting the 
zeroth-order contribution (we focus our attention on the non-local term appearing in the expansion) after some 
straightforward rearrangement we arrive at
\begin{equation}
 \mathbf{S}_{1}\cdot\mathbf{S}_{2} \simeq -\frac{d^2}{4}\int_{\Omega}d^3r\:n(\mathbf{r}) \nabla\mathbf{s}(\mathbf{r})\cdot\nabla\mathbf{s}(\mathbf{r})\:,
\end{equation}
which in turn becomes
\begin{align}
 \mathbf{S}_{1}\cdot\mathbf{S}_{2} & \simeq \frac{d^2}{4}\int_{\Omega}d^3r \bigg[-\:\nabla\cdot
 \big(n(\mathbf{r})\mathbf{s}(\mathbf{r})\cdot\nabla\mathbf{s}(\mathbf{r})\big) + \nonumber \\
 & + \mathbf{m}(\mathbf{r})\cdot\Big(\frac{\nabla n(\mathbf{r})\cdot\nabla\mathbf{s}(\mathbf{r})}{n(\mathbf{r})} + \nabla^{2}\mathbf{s}(\mathbf{r})\Big) \bigg]\:.
\end{align}
Finally, by considering a sufficiently large integration volume, the use of the divergence theorem allows to neglect all 
the boundary terms with consequent final expression
\begin{equation}
 \mathbf{S}_{1}\cdot\mathbf{S}_{2} \simeq \frac{d^2}{4}\int_{\Omega}d^3r\,\mathbf{m}(\mathbf{r})\cdot\Big[\frac{\nabla n(\mathbf{r})\cdot\nabla\mathbf{s}
 (\mathbf{r})}{n(\mathbf{r})} + \nabla^{2}\mathbf{s}(\mathbf{r})\Big]\:,
\end{equation}
which remarkable resembles the result in Eq.~(\ref{Eq:Bkin}) for the kinetic magnetic field. We can therefore tentatively interpret 
$\mathbf{B}_\mathrm{kin}(\mathbf{r}, t)$ as an effective mean-field internal magnetic field, which plays a r\^ole in coupling the spins 
at different locations in the system in the spirit of the Heisenberg spin-spin interaction.

\section{Analyzing spin dynamics from TDSDFT simulations in F\lowercase{e}$_6$ cluster} \label{res}

Here we present the results of TDSDFT calculations, performed with the Octopus code~\cite{Octop1}, where we simulate 
the ultrafast demagnetization process in iron-based ferromagnetic systems. In all those, at time $t=0$ the system is in its 
ground state. Then we apply an intense electric field pulse with a duration of less than $10$~fs, which initiates the dynamics. 
The pseudo-potentials for F\lowercase{e} used in the calculations are fully relativistic, norm-conserving and are generated using a 
Multi-Reference-Pseudo-Potential (MRPP) scheme \cite{MRPP} at the level implemented in APE~\cite{APE1, APE2}, 
which takes directly into account the semi-core states. For the XC functional we employ the ALSDA with parameterization 
from Perdew and Wang~\cite{Wang92}. Our simulations then consist in evolving in time the KS wave functions, i.e. in solving
numerically the set of equations (\ref{TDKS}). The results are then interpreted through the magnetization continuity equation 
(\ref{eq:spincon}).

\myFig{1}{1}{true}{0}{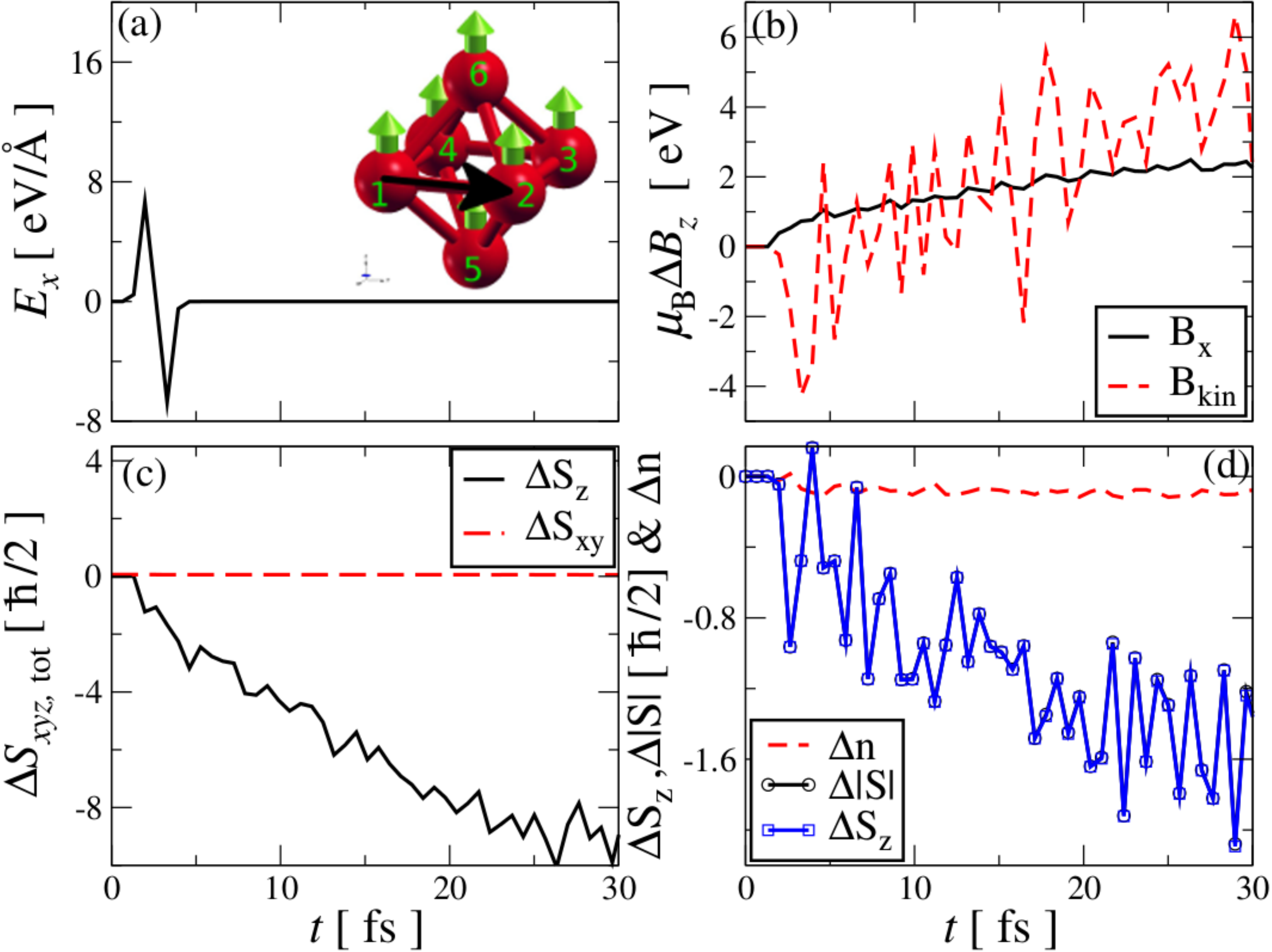}{(Color online) (a) Typical electric field pulse used to excite the F\lowercase{e}$_6$ cluster with 
the black arrow indicating the direction of the field. The fluence of this pulse is $580\,\mathrm{mJ/cm}^2$. (b) Time evolution of 
the $z$-component of $\mathbf{B}_\mathrm{kin}$ and $\mathbf{B}_\mathrm{x}$ (exchange component of the field), with
respect to their values at $t=0$ integrated over the system volume, $\mu_B\mathbf{B}_{\mathrm{tot}}(t)=\mu_B\sum_{\mathrm{I}}\mathbf{B}_{\mathrm{I}}(t)$. (c) Time evolution of the variation of the total magnetization 
$\Delta S_z^{\mathrm{tot}}(t)=\sum_{\mathrm{I}}\Delta S_z^{\mathrm{I}}(t)$ with respect to its initial value. (d) Time evolution on 
atomic site $6$ of the magnetization variation along $z$ and of the electron density variation with respect to its value at $t=0$ 
integrated inside a sphere of radius $R = \SI{0.9}{\angstrom}$. }{fig:01}

In Fig.~\ref{fig:01} the extracted magnetization dynamics of a F\lowercase{e}$_6$ magnetic cluster is presented. We use the 
LDA ground-state geometry of F\lowercase{e}$_6$ as extracted from Ref.~[\onlinecite{Dieguez},\onlinecite{gutsev}] for which 
we reproduce the reported therein spin state $S = 20~\hbar/2$. 
The nuclei are kept stationary during the dynamics. In panel (c) we observe that the total loss of the $z$ component of the 
total magnetization, $S_z^{\mathrm{tot}}(t)$, is exactly equal to the variation in value of its module, $|\mathbf{S}^{\mathrm{tot}}|$, since the global non-collinear contribution is negligible.
This indicates that the spin is not exchanged globally between the different components of the magnetization vector, but, according to 
Eq.~(\ref{eq:Tso}), it is, at least, partially transfered into the orbital momentum of the system. We note that due to the electrostatic interactions with the nuclei and due to the interaction 
with the laser field the rotational invariance of the electronic system is broken and the total orbital momentum is not conserved.

In Fig.~\ref{fig:01}(b) we observe that the average kinetic magnetic field (over the entire simulation box, for $\bar{\mathcal{F}}=1$) 
is comparable in magnitude 
to the exchange component. At the same time, $\mathbf{B}_{\mathrm{kin},z}^{\mathrm{tot}}$ shows a much more oscillatory behaviour compared to $\mathbf{B}_{\mathrm{x},z}^{\mathrm{tot}}$. 
In particular, While $\mathbf{B}_{\mathrm{x},z}^{\mathrm{tot}}$ evolves smoothly in time following the action of the optical excitation, $\mathbf{B}_{\mathrm{kin},z}^{\mathrm{tot}}$
presents an abrupt variation at the on-set of the electrical pulse. This is due to the fact that the laser pulse directly excites currents, through the term $-\nabla\cdot\mathcal{D}(\mathbf{r},t)$, 
which, in turn, produces a modification of the gradients of the charge/spin density, even on a global scale since they are not conserved. Thus we observe huge variations of 
$B_{\mathrm{kin},z}^{\mathrm{tot}}$. $B_{\mathrm{x},z}$ can also oscillate very strongly locally, following the temporal variation of the densities, but when we measure $B_{\mathrm{x},z}^{\mathrm{tot}}$ these 
oscillations are averaged out given that the densities are approximately conserved over the entire simulation box.
During the action of the pulse we see a tendency of the two fields to compensate each other, an effect strongly resembling the Lenz law. After the pulse, $\mathbf{B}_\mathrm{kin}$ continues to 
oscillate dramatically with its average value that slowly increases. In contrast $\mathbf{B}_{\mathrm{x},z}^{\mathrm{tot}}$ decreases (in absolute value) due the net dissipation of spin angular momentum. 

Moving from an analysis of global quantities to probing locally the spin dynamics, in Fig.~\ref{fig:01}(d) we compare the magnetization and 
the electron density around the atomic site $6$ at the tip of the cluster (see inset of Fig.~\ref{fig:01}(a) for the numbering labels of all the cluster atoms). 
We define local magnetization and charge associated to the particular atomic site $\mathrm{I}$ as
\begin{equation}
 \mathbf{S}^{\mathrm{I}}(t) = \int_{\mathcal{S}^{\mathrm{I}}_{\mathrm{R}}} d^3r\; \mathbf{m}(\mathbf{r},t)\:, \quad Q^{\mathrm{I}}(t) = \int_{\mathcal{S}^{\mathrm{I}}_{\mathrm{R}}} d^3r\; n(\mathbf{r},t)\:,
\end{equation}
where the integration volume $\mathcal{S}^{\mathrm{I}}_{\mathrm{R}}$ is a sphere of radius $\mathrm{R}$ centered at site $\mathrm{I}$. Our results show that the loss of 
$S_z^{6}$ is not taking place just during the action of the external pulse, but it is rather distributed over the entire time evolution. This 
suggests that the spin-sink mechanism is not directly related to the coupling of the system to the laser field, but is rather intrinsic to
the electron dynamics following the pulse. Furthermore, close to the atomic site, the temporal variation of the charge, $Q^6$, is much 
smaller in magnitude and smoother than that of $S_z^{6}$. In addition for long times $Q^{6}$ settles close to an average value, while 
$S_z^6$ continues to decrease. Hence the long-term spin dynamics is not the result of a net charge displacement from the region close 
to the ions to the interstitial space. These observations are valid for all the atomic sites in the cluster.

If we now consider the continuity equation for the electron density (see the Appendix \ref{app} for further explanations)
\begin{equation}\label{eq:chargecon}
 \frac{D}{Dt}n(\mathbf{r},t) = -n(\mathbf{r},t) \nabla\cdot\mathbf{v}(\mathbf{r},t)\:,
\end{equation}
where $\frac{D}{Dt}n(\mathbf{r},t)$ is the material derivative of the electron density
\begin{equation}
 \frac{D}{Dt}n(\mathbf{r}, t) = \Big(\frac{d}{dt} + \mathbf{v}\cdot\nabla \Big)n(\mathbf{r}, t)\:.
\end{equation}
From Fig.~\ref{fig:01}(d) we observe that during the action of the pulse the density variation in the vicinity of the atoms appears to be very small compared to the
magnetization variation. We can therefore safely assume that in this spatial region $\dot{n}(\mathbf{r}, t)\simeq 0$, with at the same time $n(\mathbf{r}, t)\neq 0$. 
From these considerations we deduce that $\mathbf{v}(\mathbf{r},t)\simeq 0$ is a reasonably good approximation for the velocity field in the vicinity of the atoms (this does not
imply that the velocity field is exactly zero, but only that its effect on the spin dynamics in this particular case is negligible).
The same argument is valid also for the state resolved density $n_j(\mathbf{r},t)$, given that $\dot{n}(\mathbf{r},t)=\sum_{j\in\mathrm{occ.}}\dot{n}_j(\mathbf{r},t)$, the contribution
of the local time derivative of the Kohn-Sham state density can be neglected. 
By applying the latter into Eq.~(\ref{eq:spincon}) we finally obtain a relation that could be considered approximately valid in this spatial region of the simulation box,
\begin{equation}\label{Eq:effspcon}
\frac{d}{dt}{\mathbf{m}}(\mathbf{r}, t) \simeq -\nabla\cdot\mathcal{D} + \mu_B\mathbf{m}\times\mathbf{B}_{\mathrm{kin}} + \mathbf{T}_\mathrm{SO}\:,
\end{equation}
where the contribution to the spin dynamics due to the velocity field term has been neglected.
Note that here we have also used the condition $\mathbf{m}(\mathbf{r}, t)\times \mathbf{B}_\mathrm{xc}(\mathbf{r}, t) = 0$, that is consequential to
the local density approximation. In addition, the decay of $\mathbf{B}_{\mathrm{xc}}$, during the evolution, is not so relevant to justify a dynamical modification of the gap
between up and down spin states. 

\myFig{1}{1}{true}{0}{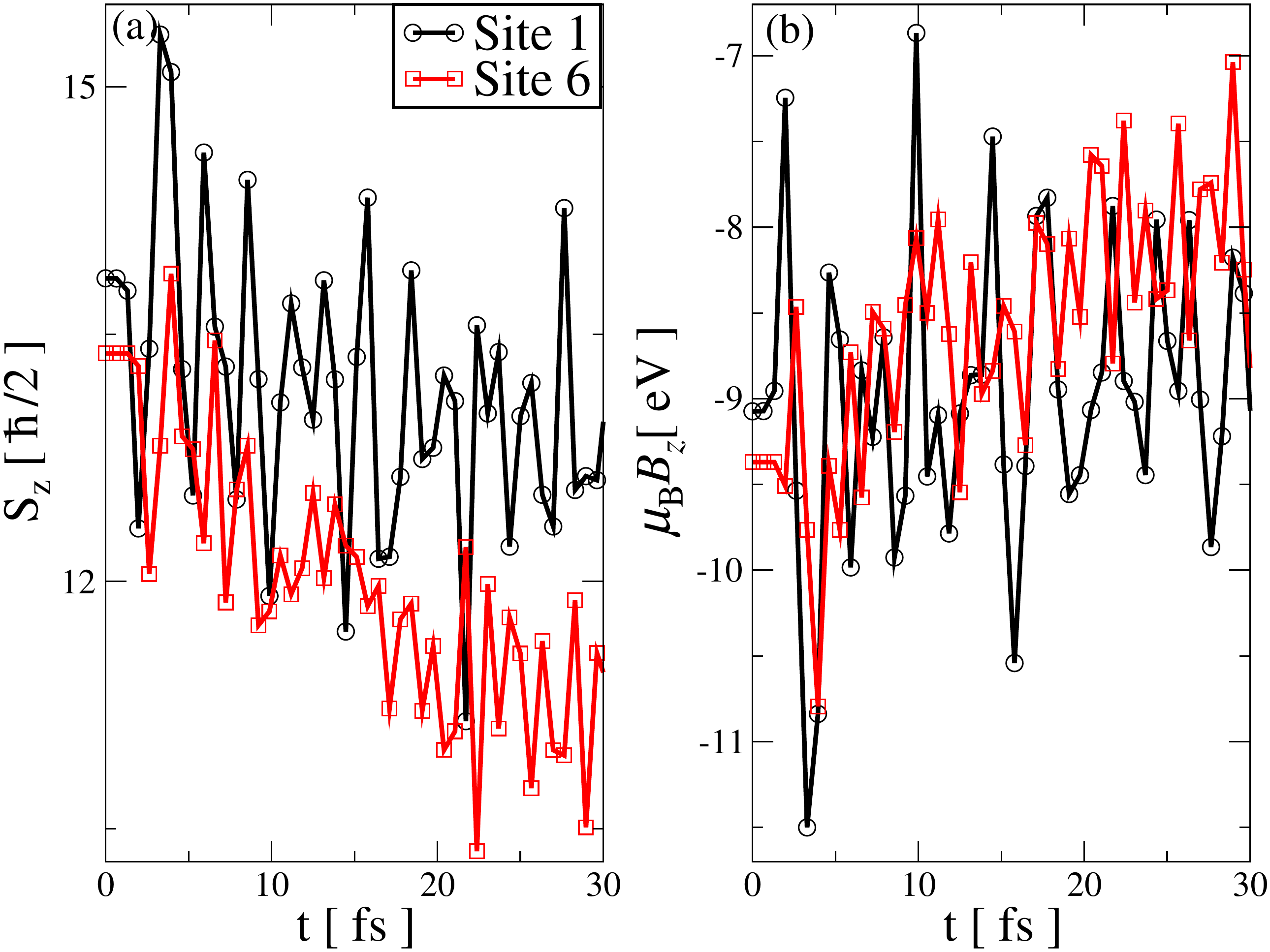}{(Color online) Local spin dynamics of the F\lowercase{e}$_6$ cluster: (a) Time evolution of the magnetization $S_z^{\mathrm{I}}(t)$ around the 
atomic centers; (b) time evolution of the $z$ component of $\mathbf{B}_{\mathrm{kin}}^{\mathrm{I}}(t)$. All the quantities are integrated inside a sphere of radius $R = \SI{0.9}{\angstrom}$ centered 
on the two atomic sites, where we have used $B_{z}^{\mathrm{I}}(t) = \int_{\mathcal{S}_{\mathrm{R}}^{\mathrm{I}}}d^3r\:B_{z}(\mathbf{r},t)$. }{fig:02}

In Fig.~\ref{fig:02} we compare the behaviour of the kinetic field and of the local magnetization at two atomic sites, respectively $1$ 
(one of the atoms in the base plane of the bi-pyramid) and $6$ (an atom at one of the apexes). It can be seen from panel (a) that 
these two sites present different rates of demagnetization. In particular, at site $6$ the spin decay is considerably more prominent 
with respect to that observed at site $1$. In contrast the fluctuations in $S_z^{\mathrm{I}}$ are significantly more pronounced for site $1$ than for
site $6$. This can be understood from the fact that we have chosen here an electric pulse with polarization vector in the basal plane of
the bi-pyramid. As such, the charge fluctuations for the atoms in the basal plane are expected to be much larger than those of the 
apical atoms. Finally, we note that $B_{\mathrm{kin},z}^{\mathrm{I}}(t)$ follows similar qualitative trends as $S_z^\mathrm{I}(t)$ 
[see Fig.~\ref{fig:02}(b)]. In fact, the average change following the excitation pulse is larger for site $6$ (the one experiencing the
larger demagnetization), but the fluctuations are more pronounced for site $1$ (the one experiencing the larger fluctuations in $S_z^{\mathrm{I}}(t)$).

The correlation between the kinetic field and the magnetization loss is also rather evident in Fig.~\ref{fig:03}. There the time-averaged 
variations in the $x$-component of the two fields $\mathbf{m}\times\mathbf{B}_{\mathrm{kin}}$ and $\dot{\mathbf{s}}(\mathbf{r},t)$ are clearly 
comparable in magnitude and localized over the same regions of the simulation box. This demonstrates that the kinetic field can be considered as
the main force driving the non-collinearity during the spin evolution. The fact that the contrast is stronger at the apex atoms (``hot spots'' for 
demagnetization) agrees with Fig.~\ref{fig:02}(a), while the dipole-type patterns indicate how the longitudinal spin decays preserving 
global collinearity. The correlation between the $z$ components of $\mathbf{m}\times\mathbf{B}_{\mathrm{kin}}$ and 
$\dot{\mathbf{s}}(\mathbf{r}, t)$ is not as evident as that for the transverse component $x$. This is due to the fact that the $x$ 
and $y$ components of the field are much smaller compared to the $z$ one. Furthermore, the contribution to the spin dynamics along $z$ of the SO coupling, together 
with the internal dissipative term due to the spin currents, cannot be neglected.

\myFig{1}{1}{true}{0}{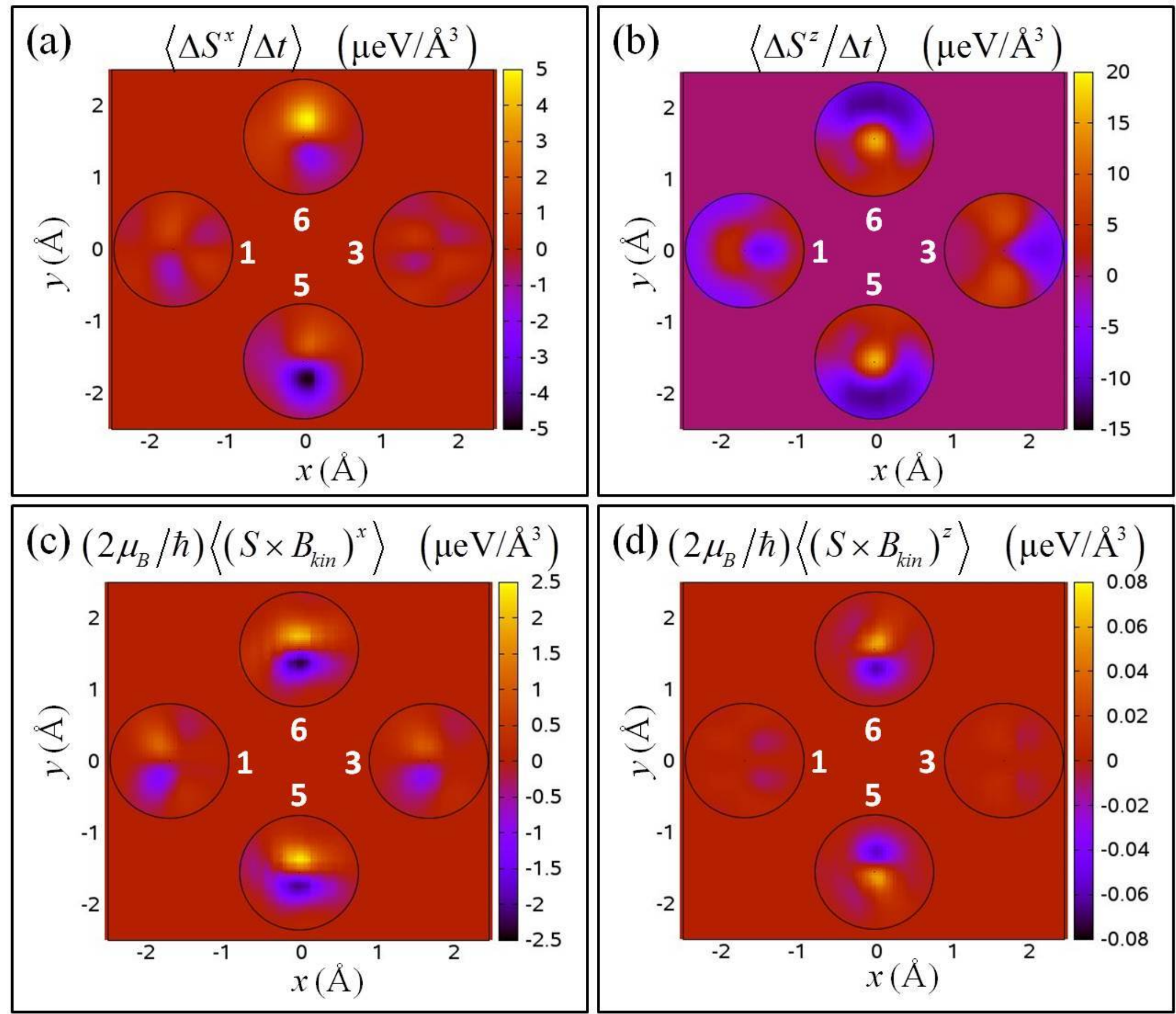}{(Color online) Contour plots of the time- and space-averaged (in direction perpendicular to the plane spanned 
by atoms $1$, $3$, $5$ and $6$, as indicated on the plot) observables evaluated only within spheres of radius $R = \SI{1.0}{\angstrom}$ around each atom: 
(a) and (b) the temporal variation of the spin density $\Delta \mathbf{s}^{(x,z)}(\mathbf{r},t)/\Delta t$ for $\Delta t = 0.1~fs$; (c) and (d) the $x$ and $z$ components of 
the second term on the right-hand side of Eq.~(\ref{Eq:effspcon}). }{fig:03}

In order to quantify the local non-collinearity we examine the evolution of the misalignment angle, $\theta$, between the $z$-axis and the 
direction of the magnetic fields (averaged over spheres). It can be seen in Fig.~\ref{fig:04} that at site $1$ the averaged kinetic field and 
the local spin deflect very little from the quantization axis and remain rather parallel to each other. The angle that $\mathbf{B}_{\mathrm{kin}}^1(t)$ forms with the magnetization 
direction (the $\mathbf{m}^1$ direction) is substantially negligible. Instead, at site $6$, $\mathbf{B}_{\mathrm{kin}}^6(t)$ shows a significant 
deflection from the $z$-axis after the first $5$~fs of the evolution and so does the spin, without the two being parallel to each other. It is important to notice that the angle
between magnetization $\mathbf{m}$ and $\mathbf{B}_{\mathrm{kin}}$ starts to grow only after the action of the pulse.
These results for atom $1$ and $6$ are representative for all the other sites in the base plane or outside of it, respectively. The sites located in the plane, 
where $\mathbf{B}_{\mathrm{kin}}^{\mathrm{I}}$ is mostly collinear, lose less magnetization with respect to the ones at the apices where, instead, 
the kinetic field shows a significant deflection from the magnetization axis and provides additional torque driving further demagnetization. 

Analogous conclusions arise from the introduction of the concept of parallel transport, commonly used in differential geometry. This requires a 
proper definition of the covariant derivative obtained by comparing $\mathbf{s}(\mathbf{r}+d\mathbf{r})$ not with $\mathbf{s}(\mathbf{r})$, 
but with the value that the spin vector would have if it was translated from $\mathbf{r}$ to $\mathbf{r}+d\mathbf{r}$ while keeping the axes 
in the isospin space fixed,
\begin{equation}\label{eq:connect}
 D_{i}\mathbf{s}(\mathbf{r}, t) = d_{i}\mathbf{s}(\mathbf{r}, t) + \boldsymbol{\mathcal{A}}_i(\mathbf{r}, t)\times \mathbf{s}(\mathbf{r}, t)\:.
\end{equation}
The connection field $\boldsymbol{\mathcal{A}}(\mathbf{r},t)$ provides a measure of the amount of non collinearity accumulated in the 
translation of the spin vector from $\mathbf{r}$ to $\mathbf{r}+d\mathbf{r}$

\myFig{1}{1}{true}{-90}{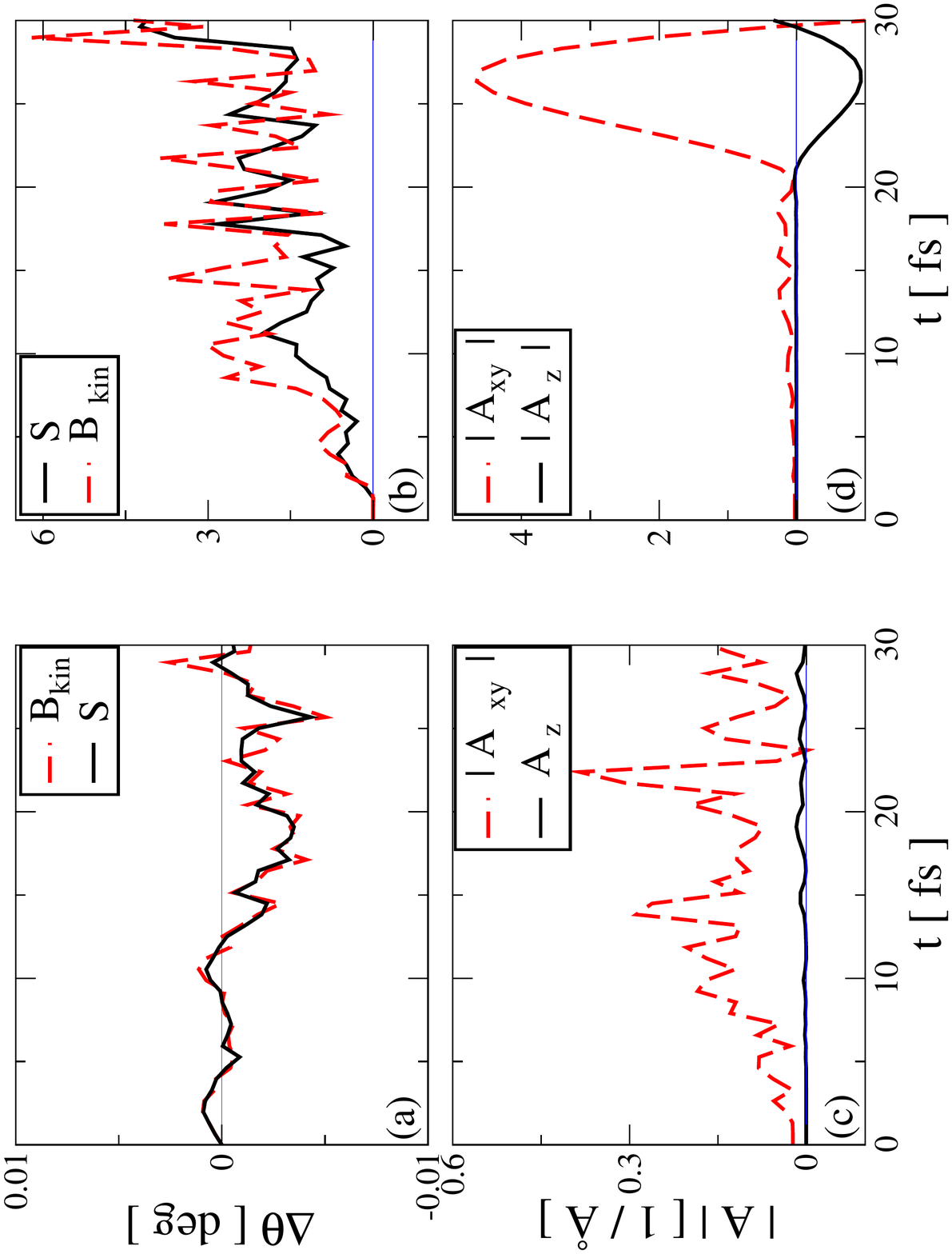}{(Color online) Evolution of the spin non-collinearity in the F\lowercase{e}$_6$ cluster. (a) $\Delta\theta = \theta(t) - \theta(0)$ at site $1$, for 
the $\mathbf{B}_{\mathrm{xc}}$ [or $\mathbf{S}(t)$] direction (black curve) and the $\mathbf{B}_{\mathrm{kin}}$ direction (red dashed curve). (b) The same quantities of panel (a) but calculated 
at the atomic site $6$. (c) $\sqrt{\bar{\mathcal{A}}_1^2+\bar{\mathcal{A}}_2^2}$ where $\bar{\boldsymbol{\mathcal{A}}}=\sqrt{\sum_{i=1}^3\boldsymbol{\mathcal{A}}_i^2}$ and $\boldsymbol{\mathcal{A}}_i$ 
is introduced in Eq.~(\ref{eq:connect}), compared to $\bar{\mathcal{A}}_3$ at the atomic site $1$. (d) The same quantities of panel (c) but calculated at the atomic site $6$. 
The fields are measured within a sphere of radius $R = \SI{0.8}{\angstrom}$ centered on the atom center.}{fig:04}

By using the previous expression to rewrite the first and second order spatial derivatives, the kinetic field of Eq.~(\ref{Eq:Bkin}) can be divided in two components
\begin{equation}
\mathbf{B}_{\mathrm{kin}}(\mathbf{r},t) = \mathbf{B}^{0}_{\mathrm{kin}}(\mathbf{r}, t) + \delta\mathbf{B}_{\mathrm{kin}}(\mathbf{r}, t)\:.
\end{equation}
Here we have introduced
\begin{equation}
 \mathbf{B}^{0}_{\mathrm{kin}}(\mathbf{r}, t) = \frac{1}{\bar{\mathcal{F}}e}\Big[ \frac{\nabla n(\mathbf{r}, t)}{n(\mathbf{r}, t)}\cdot D\mathbf{s}(\mathbf{r}, t) + D^{2}\mathbf{s}(\mathbf{r}, t) \Big]\:,
\end{equation}
which has no effects on the dynamics, having locally the same direction of the spin vector by construction, and
\begin{align}
\delta\mathbf{B}_{\mathrm{kin}}(\mathbf{r},t) & = \frac{1}{\bar{\mathcal{F}}e}\sum_{i=1}^{3}\Big[ \frac{d_{i}n}{n}\big(-\mathbf{s}^{2}\boldsymbol{\mathcal{A}}_i + (\mathbf{s}\cdot\boldsymbol{\mathcal{A}}_i)\mathbf{s}\big) - \nonumber \\
 &- 2\big(\mathbf{s}^{2} d_i\boldsymbol{\mathcal{A}}_i - \mathbf{s}(\mathbf{s}\cdot d_{i}\boldsymbol{\mathcal{A}}_i)\big) - 2\big(\boldsymbol{\mathcal{A}}_i(\mathbf{s}\cdot D_{i}\mathbf{s}) - \nonumber \\
 &- D_{i}\mathbf{s}(\mathbf{s}\cdot\boldsymbol{\mathcal{A}}_i)\big) + 4(\mathbf{s}\times\boldsymbol{\mathcal{A}}_i)(\mathbf{s}\cdot\boldsymbol{\mathcal{A}}_i)\Big]\:.
\end{align}
According to Fig.~\ref{fig:04}, in the case of F\lowercase{e}$_6$, the direction in the isospin space of the connection tensor $\boldsymbol{\mathcal{A}}_i$ for every $i$ 
component can be considered in first approximation orthogonal to the direction of the spin vector $\mathbf{s}(\mathbf{r},t)$, since its component 
along $z$ is considerably smaller than the components along $x$ and $y$. From this we could assume $\mathbf{s}\cdot\boldsymbol{\mathcal{A}}_i\simeq 0$ and we obtain the following simplified expression 
for $\delta\mathbf{B}_{\mathrm{kin}}(\mathbf{r},t)$
\begin{equation}
 \delta\mathbf{B}_{\mathrm{kin}}(\mathbf{r},t) = \frac{1}{\bar{\mathcal{F}}e}\sum_{i=1}^{3}\bigg[ \Big(-\frac{d_{i} n}{n}\mathbf{s}^{2} - 2\mathbf{s}\cdot D_{i}\mathbf{s}\Big)\cdot\boldsymbol{\mathcal{A}}_i - 2\mathbf{s}^{2}d_{i}\boldsymbol{\mathcal{A}}_i\bigg]\:.
\end{equation}
This represents the part of $\mathbf{B}_\mathrm{kin}$ which gives rise to a non-zero torque in Eq.~(\ref{Eq:effspcon}). 

\section{Directionality of the demagnetization in F\lowercase{e}$_6$ cluster} \label{Edir}

In the previous sections we have revisited the concept of kinetic field, its derivation within DFT and its properties as 
a major source of torque for the spin dynamics within ALSDA. We have provided supporting evidence for the latter from 
TDSDFT calculations of the ultrafast demagnetizing F\lowercase{e}$_6$ cluster under the effect of a single fs electric field pulse. Despite the 
conceptual clarity of $\mathbf{B}_{\mathrm{kin}}$ as an instrumental object, very little useful physical intuition can be drawn from 
its definition in Eq.~(\ref{Eq:Bkin}). Clearly, it is an intrinsic dynamic field that depends on the spin texture and its response to the 
external stimuli. It also feeds back into the dynamics of this same spin density, clearly a non-linear process. In this section we seek 
to extend the evidential base for the connection between the torque due to $\mathbf{B}_{\mathrm{kin}}$ and the rate of demagnetization. 
Together with that we report a situation, where the direction of the polarization vector of the electric field of the laser pulse alone 
has a significant effect on the demagnetization of a material (the F\lowercase{e}$_6$ cluster).

\myFig{1}{1}{true}{0}{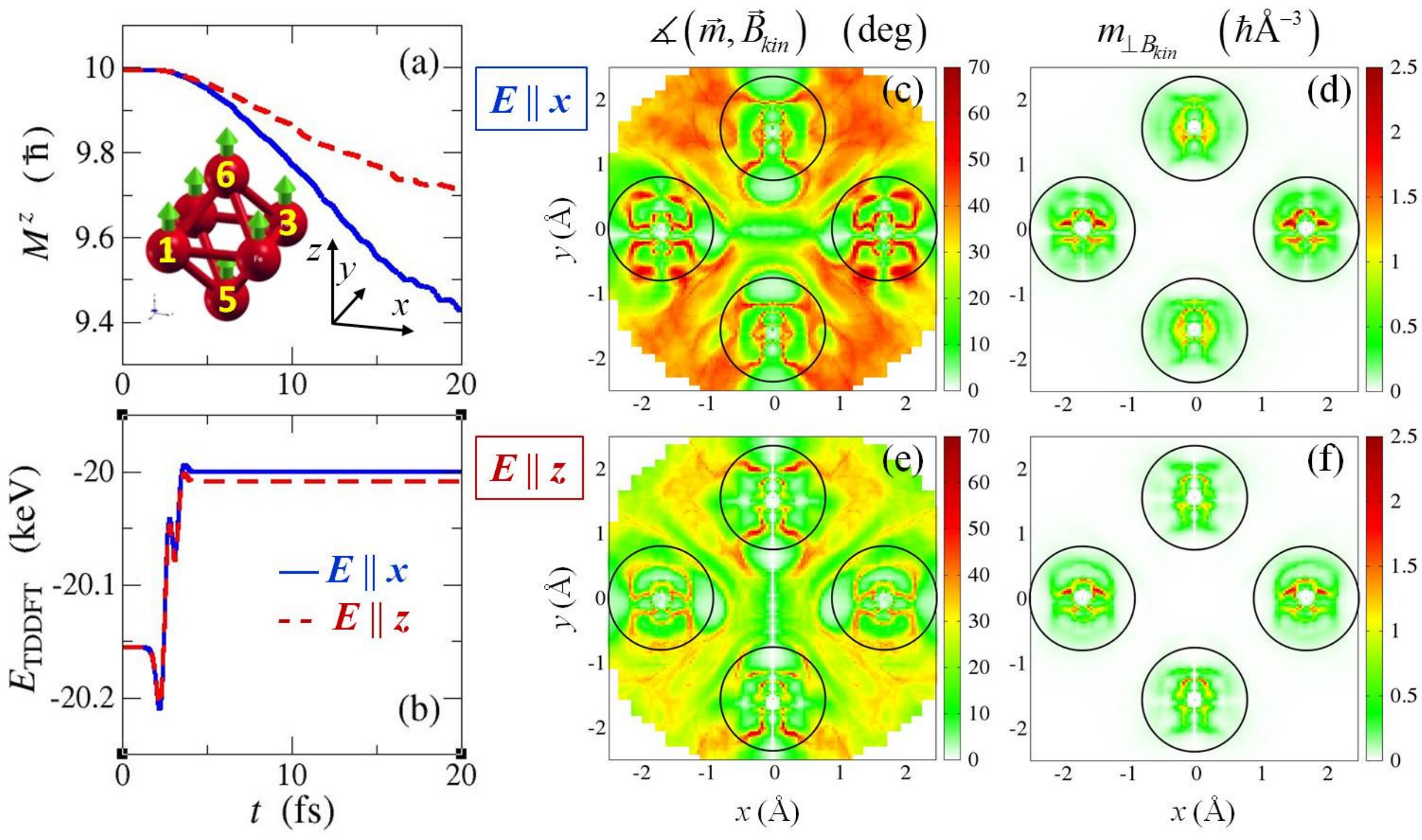}{(Color online) Global (a) spin and (b) energy variation in Fe$_6$ for two different excitations differing 
only by the direction of polarization of the electric field pulse. Cartoon of the cluster with labels of the relevant atoms and a reference 
frame are depicted as insets. The contour plots represent the distribution of the angle (c, e) between $\mathbf{m}$ and $\mathbf{B}_{\mathrm{kin}}$ 
in a plane through atoms $1$, $3$, $5$ and $6$ (as in Fig. \ref{fig:03}) and the perpendicular component of $\mathbf{m}$ with respect to 
$\mathbf{B}_{\mathrm{kin}}$ (d, f) averaged over the time of the simulation, for the two different excitations $\mathbf{E} || \mathbf{x}$ and 
$\mathbf{E} || \mathbf{z}$, respectively the top and bottom panels.}{fig:05}

Figure~\ref{fig:05} shows a comparison between two simulations differing only by the direction (but notably not the magnitude) of the 
electric field applied. In one case this is in the $\mathbf{x}$-direction, which is oriented along the slightly longer side of the 
base of the bi-pyramid \cite{Dieguez}, and in the other simulation it is along $\mathbf{z}$, the direction connecting the two apex 
atoms $5$ and $6$. The case $\mathbf{E}||\mathbf{x}$ shows nearly 3 times faster demagnetization compared to the $\mathbf{E}||\mathbf{z}$ 
one. Evidently, in the former situation more energy is absorbed by the cluster (an excess of about $\SI{8.6}{eV}$), we show a comparison of the 
total energy shift due to the pulse in Fig.~\ref{fig:05}(b).

From panel (c) we observe that the amount of non-collinearity enclosed in a relatively small radius around the apex atoms does not
significatively change in the two cases, suggesting that the intra-site non-collinear component of the spin vector is already present
in the ground-state configuration. What really differs is the amount of inter-sites non collinearity concentrated in the out-of-plane
region. In the $\mathbf{E}||\mathbf{x}$ case, if we examine the angle between $\mathbf{B}_{\mathrm{kin}}(\mathbf{r},t)$ and $\mathbf{m}(\mathbf{r},t)$
averaged in time over the entire simulation along one particular cross-section plane (vertical through the base diagonal of the cluster as
depicted in the inset), one can observe, in particular in the out-of-plane interstitial regions, a significant larger amount of non-collinearity,
of the spin vector density, accumulating in the case of faster demagnetization.

Notably, there is a change in the symmetry between panels (c) and (e) in Fig.~\ref{fig:05} -- in both cases no significant spin non-collinearity arises in the plane parallel 
to the field connecting the atomic centers. It is, however, important to notice that the amount of intra-site spin non-collinearity for the
in-plane atoms is strongly dependent on the polarization direction of the applied laser field. In fact, in the case of $\mathbf{E}||\mathbf{x}$ the temporal
averaged angle between $\mathbf{s}$ and $\mathbf{B}_{\mathrm{kin}}$ appears much higher with respect to the one computed in the $\mathbf{E}||\mathbf{z}$ case. This
quantity is mostly averaged out by the integration procedure but it is clearly visible in the panel (c) of the figure.

Although $\mathbf{B}_{\mathrm{kin}}$ is not the only torque generator and the SO contribution is significant too, the former plays a r\^ole in 
deflecting the spins in the system in a manner that correlates with the rate of global demagnetization. Importantly, our simulations 
clearly show that the demagnetization process is very anisotropic and particular directions of the exciting electric field may enhance the rate of demagnetization
(the study of these effects is beyond the scope of this paper and it will be explored in more details in a coming publication).

\section{Demagnetization of \lowercase{bcc} F\lowercase{e}} \label{bccFe}

Finally, we present results of analogous simulations in bulk materials, namely in bcc F\lowercase{e}, with the aim of demonstrating the qualitative 
universal r\^ole played by $\mathbf{B}_{\mathrm{\mathrm{kin}}}$ in the ultrafast demagnetization process. We consider bcc F\lowercase{e} in its ferromagnetic 
phase with total spin in the unit cell $S = 14.97\hbar/2$ and with 2 atoms in it.

\myFig{1}{1}{true}{0}{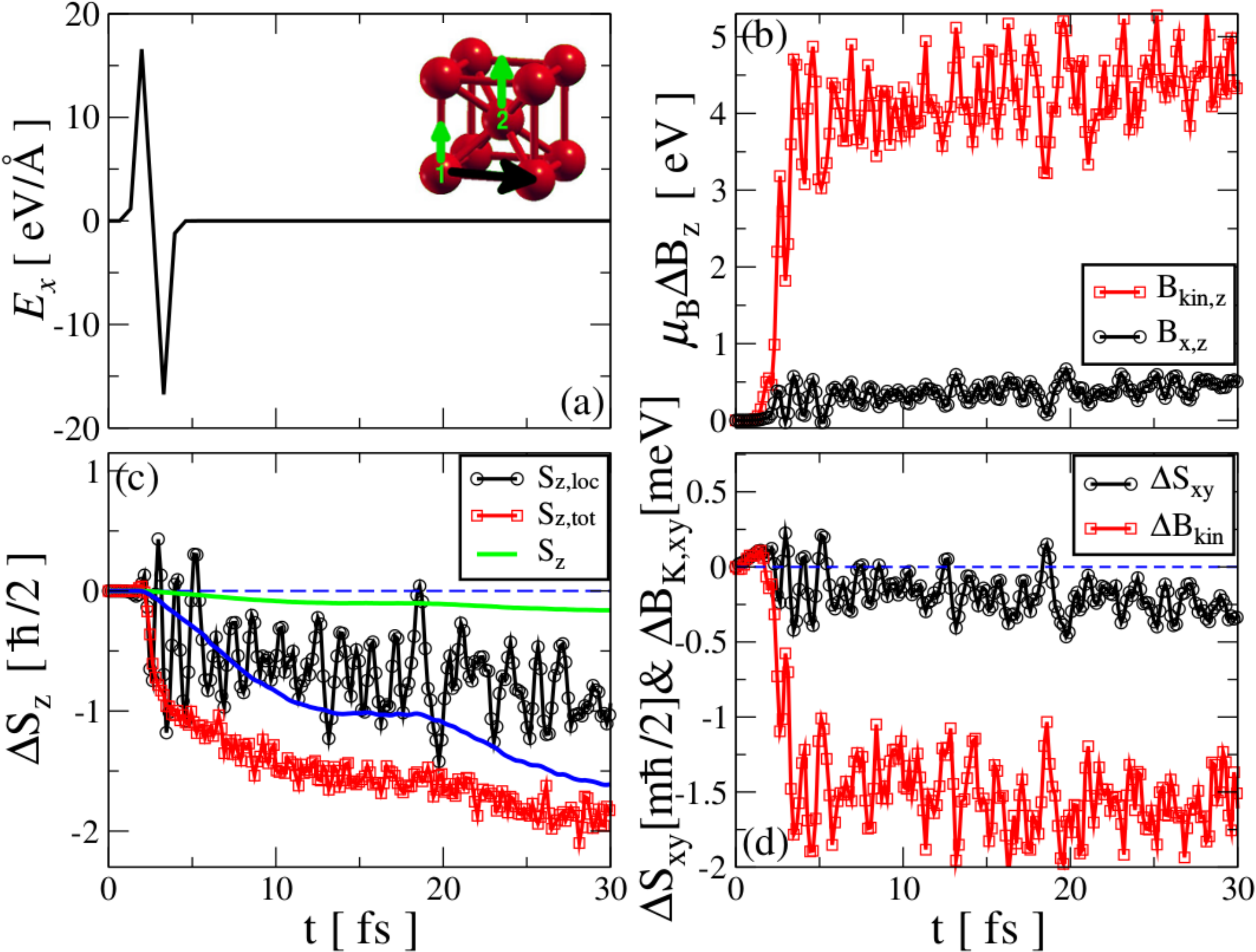}{(Color online) Demagnetization of bcc F\lowercase{e}: (a) applied external electric field, (b) local value of $\Delta B_{\mathrm{kin},z}^{\mathrm{I}}$ and
$\Delta B_{\mathrm{x},z}^{\mathrm{I}}$ around atom $1$; (c) comparison between the value of the local magnetization $\Delta S_{z,\mathrm{I}}(t)$ around atom $1$ (black curve), the total 
magnetization integrated around the two sites (red curve), $\Delta S_{z,\mathrm{I}}(t)$, and the total magnetization integrated inside the unit cell $\Delta S_z(t)$ (green curve); (d) local 
value of $\sum_{\mathrm{I}}\Delta B_{\mathrm{kin},xy}^{\mathrm{I}}(t)$ (red curve) and of $\Delta S_{xy}^{\mathrm{I}}(t)$, non collinear magnetization component for atom $1$. All the local quantities 
are calculated inside a sphere of radius $R = \SI{0.8}{\angstrom}$ centered on site $1$. }{fig:06}

We employ a lattice parameter $a = \SI{2.9}{\angstrom}$, with a $4\times4\times4$ $k$-points grid. In Fig.~\ref{fig:06}(c) we show the 
demagnetization rate of the single unit cell after it has been excited with the electric field pulse [panel (a)]. The green curve represents the
demagnetization computed inside the unit cell and resembles in shape the sum of the magnetization variation, $\sum_{\mathrm{I}}\Delta S_{z,\mathrm{I}}(t)$, calculated
in the vicinity of the two F\lowercase{e} atoms, even if it is different in magnitude. This suggests that a large amount of spin is driven outside from the atomic
integration region during the evolution.
Similarly to the cluster, the dynamics of the onsite magnetization can be described in terms of a two-step process with an initial fast decay during the action of the 
external pulse, followed by a slower and noisy decrease in magnitude. The first fast decay may be attributed to the effect of the SO enhanced by the collapse of the
effective field $\mathbf{B}_{\mathrm{eff}}$ following the action of the laser pulse. In Fig.~\ref{fig:06}(b) the collapse after the first $5~fs$ of the $z$ component of
the effective field is quite clear, even if it appears to be more pronounced for the kinetic field $B_{\mathrm{kin},z}^{\mathrm{I}}(t)$ with respect to the exchange field $B_{\mathrm{x},z}^{\mathrm{I}}(t)$.
Similarly to the case of F\lowercase{e}$_6$ the role played by $\nabla\cdot\mathcal{D}(\mathbf{r},t)$ is dominant during the action of the pulse, but after this initial
phase the dynamics is dominated by intra-band transitions and the interplay between the spin-orbit coupling and the effective field $\mathbf{B}_{\mathrm{eff}}$ becomes dominant.

Fig.~\ref{fig:06}(d) shows the evolution of the non-collinearity of the spin vector, $\sum_{\mathrm{I}}S_{xy}^{\mathrm{I}}(t)$, and of the kinetic field, $\sum_{\mathrm{I}}B_{\mathrm{kin},xy}^{\mathrm{I}}(t)$.
The level of correlation among the two quantities confirms the importance of the kinetic field in the evolution of the spin non-collinearity. The long tail of spin dissipation
may be explained in terms of intra-band spin-up/spin-down transitions through an Elliott-Yafet type of mechanism triggered by the scattering with the effective field $\mathbf{B}_{\mathrm{eff}}$,
\begin{equation}
\mathcal{A}_{i\rightarrow f} = \bra{\Psi_{n,\mathbf{k}_1}}\hat{\boldsymbol{\sigma}}\cdot\mathbf{B}_{\mathrm{eff}}\ket{\Psi_{n,\mathbf{k}_2}}.
\end{equation}
$\mathcal{A}_{i\rightarrow f}$ represents the transition amplitude between two states with different $\mathbf{k}$ vector and in presence of SO with different mixing of up and down
spin components.

\section{Conclusions} \label{concl}
In conclusion, we remark the central result of our work, namely that the equation of motion for the spin dynamics within 
the ALSDA of TDSDFT [see Eq.~(\ref{Eq:SpinCon})] can be rewritten in the form of Eq.~(\ref{eq:spincon}), by using a 
formalism borrowed from magneto-hydrodynamics. Subsequently we have analyzed the properties of the so-called kinetic 
magnetic field $\mathbf{B}_\mathrm{kin}$ and its r\^ole in the ultrafast demagnetization process in two different systems: 
a ferromagnetic Fe$_6$ cluster and bulk bcc Fe. The r\^ole of this field is particularly significant for processes far from 
equilibrium, such as the ultrafast demagnetization observed in transition metals.

In both the systems studied the spin dynamics is the result of the interplay between the SO coupling and 
$\mathbf{B}_{\mathrm{kin}}(\mathbf{r},t)$, which, in general, is strongly coupled to the external pulse and highly 
non-uniform in space. We have shown that the spin loss locally correlates with $\mathbf{B}_{\mathrm{kin}}(\mathbf{r}, t)$. 
Through the concept of parallel transport and the definition of a connection tensor field $\boldsymbol{\mathcal{A}}_{i}$, 
we have gained further insight into the evolution of the spin texture. As $\boldsymbol{\mathcal{A}}_{i}$ describes the 
degree of spin rotation per infinitesimal spatial translation, it also provides a measure for the misalignment between the 
kinetic field and the spin texture. The regions with higher $\Norm{\boldsymbol{\mathcal{A}}}$ correspond to stronger 
local demagnetization. 

Finally, the effect of the direction of the polarization vector of the electric field pulse has been studied for Fe$_6$. We 
have found that clusters will demagnetize about twice as fast, if the polarization vector is in the base plane and not 
vertical (through the apex atoms). Our analysis has shown a significant increase in the non-collinearity between 
$\mathbf{B}_{\mathrm{kin}}(\mathbf{r},t)$ and the spin density in the fast demagnetizing case. Such anisotropy, due 
to the electric dipole matrix elements for the valence electrons, is likely to occur in crystalline systems as well. During 
the application of the laser pulse, the rise of spin non-collinearity may be enhanced by the particular polarization
direction of the laser pulse through the spin orbit coupling and this effect combined with the collapse of the kinetic field 
may explain the initial spin loss. However, in both F\lowercase{e}$_6$ and F\lowercase{e} bcc the magnetization loss 
is more prominent after that the laser pulse has been set to zero. During this second phase of spin dissipation we 
need to distinguish between the spin decay observed in bcc F\lowercase{e} due to intra-band transitions among 
states with different spin up/down mixing and the spin dynamics observed in correspondence of the apex atoms in 
F\lowercase{e}$_6$ that is, instead, driven by $\mathbf{B}_{\mathrm{kin}}$ and directly related to the onsite intrinsic spin 
non-collinearity near the atomic sites.  

\acknowledgments
This work has been funded by the European Commission project CRONOS (grant no. 280879) and by Science 
Foundation Ireland (grant No. 14/IA/2624). We gratefully acknowledge the DJEI/DES/SFI/HEA Irish Centre for 
High-End Computing (ICHEC) for the provision of computational facilities. We also thank Prof.~E.K.U. Gross for 
valuable discussions.

\appendix 

\section{Derivation of the hydrodynamic continuity equation} \label{app}

In this appendix we show in some detail how Eq.~(\ref{eq:spincon}) can be obtained by starting from the 
standard TDSDFT continuity equation~(\ref{Eq:SpinCon}). Here we follow the hydrodynamical formalism 
of quantum mechanics, where a single particle with spin is considered equivalent to a non-linear vector 
field. In this type of hydrodynamics the quantum effects are separated as non-linear terms and are described 
through effective quantum potentials (see Ref.~[\onlinecite{Taka55}]).

The formalism is based on the assumption that it is possible to describe the dynamical evolution of a single 
particle immersed in an external vector potential, $\mathbf{A}(\mathbf{r},t)$, through the so-called Madelung 
decomposition (see Ref.~[\onlinecite{Holl93}]) of the system wave function, in which the amplitude is translated 
into the probability density and the gradient of the phase determines the velocity field. A hydrodynamical description 
of the wave function was also obtained in Ref.~[\onlinecite{Schon54}] starting from the ordinary interpretation of
quantum mechanics and by introducing an operator for the charge density and the current density.

The formalism was also later extended to the semi-relativistic description (Pauli approximation) of a single particle 
in an external electro-magnetic field in Ref.~[\onlinecite{hydro2}]. However, while in all the previous studies the main 
objective was to derive the single-particle dynamics of the spin $1/2$ plasma, only recently the study of the collective 
dynamical properties of the quantum plasma started to attract some interests (see Ref.~[\onlinecite{hydro3}] and 
[\onlinecite{Brod11}]).

In deriving the Eq.~(\ref{eq:spincon}) for the spin density in the Kohn-Sham system we introduce also of the electron 
density, $n(\mathbf{r},t)$, and the velocity field, $\mathbf{v}(\mathbf{r},t)$. The equation of motion for the velocity field 
are not explicitly written here for the reasons explained in section~\ref{res}.
 
The electron density is written in terms of the Kohn-Sham wave functions $\psi_j^{\mathrm{KS}}(\mathbf{r},t)$ as
\begin{widetext}
\begin{equation}
 n(\mathbf{r}, t) = \sum_{j\in \mathrm{occ.}}\psi_{j}^{\mathrm{KS}}(\mathbf{r}, t)^{\dagger}\psi_{j}^{\mathrm{KS}}(\mathbf{r}, t) \:,
\end{equation}
while the spin density is
\begin{equation}
 \mathbf{s}(\mathbf{r}, t) = \frac{\sum_{j\in \mathrm{occ.}}\psi_{j}^{\mathrm{KS}}(\mathbf{r}, t)^{\dagger}\bm{\sigma}\psi_{j}^{\mathrm{KS}}(\mathbf{r}, t)}{n(\mathbf{r}, t)}\:,
\end{equation}
and the covariant velocity field appears as
\begin{equation}
 \mathbf{v}(\mathbf{r}, t) = \frac{\hbar}{2mi}\cdot\frac{\sum_{j\in \mathrm{occ.}}(\psi_{j}^{\mathrm{KS}}(\mathbf{r}, t)^{\dagger}\nabla\psi_{j}^{\mathrm{KS}} - \psi_{j}^{\mathrm{KS}}\nabla\psi_{j}^{\mathrm{KS}}(\mathbf{r}, t)^{\dagger})}{n(\mathbf{r}, t)} - \frac{e}{mc}\mathbf{A}(\mathbf{r}, t)\:.
\end{equation}
By making use of the Kohn-Sham equations (\ref{TDKS}) the charge continuity equation can be written 
straightforwardly as
\begin{equation}\label{eq:contn}
 \frac{d}{dt}n(\mathbf{r},t) = -\frac{\hbar}{2mi}\nabla\cdot\sum_{j\in\mathrm{occ.}}\big[\psi_{j}^{\mathrm{KS}}(\mathbf{r}, t)^{\dagger}(\overrightarrow{\nabla} - \overleftarrow{\nabla})\psi_{j}^{\mathrm{KS}}(\mathbf{r}, t) \big] + \frac{e}{mc}\nabla\cdot\big[ n\mathbf{A}(\mathbf{r}, t)\big]\:,
\end{equation}
while for spin it is written in terms of $n(\mathbf{r},t)$ and $\mathbf{s}(\mathbf{r},t)$ as
\begin{equation}\label{eq:SC}
\frac{d}{dt}(n\mathbf{s}) = -\frac{\hbar^{2}}{4mi}\nabla\cdot\sum_{j\in\mathrm{occ.}}\big[ \psi_{j}^{\mathrm{KS}}(\mathbf{r}, t)^{\dagger}\hat{\boldsymbol{\sigma}}\nabla\psi_{j}^{\mathrm{KS}} - \nabla\psi_{j}^{\mathrm{KS}}(\mathbf{r}, t)^{\dagger}\hat{\boldsymbol{\sigma}}\psi_{j}^{\mathrm{KS}}\big] + \frac{e}{mc}\sum_{j}\partial_{j}\big[ A^{j}n\mathbf{s}\big] + \mu_{B}n(\mathbf{s}\times\mathbf{B}_{\mathrm{xc}}) + \mathbf{T}_{\mathrm{SO}} \:.
\end{equation}
By evaluating explicitly the spatial partial derivative of the spin vector and by multiplying it with the component $s_i$ we 
obtain the following equality
\begin{equation}\label{Eq:2}
n s_{i}\partial_{l}s_{k} = s_{i}\sum_{j\in\mathrm{occ.}}\big[ \partial_{l}\psi_{j}^{\mathrm{KS}}(\mathbf{r},t)^{\dagger}\hat{\sigma}_k\psi_{j}^{\mathrm{KS}} + \psi_{j}^{\mathrm{KS}}(\mathbf{r},t)^{\dagger}\hat{\sigma}_k\partial_{l}\psi_{j}^{\mathrm{KS}}\big] -\partial_{l}n\, s_{i}\cdot s_{k}\:.
\end{equation}
We now need to focus our attention on the first term on the right-hand-side of Eq.~(\ref{Eq:2}), by using the following notation
\begin{equation}
\mathcal{F}_{ik}(\mathbf{r},t) = s_i \sum_{j\in\mathrm{occ.}}\big[ \nabla\psi_j^{\mathrm{KS}}(\mathbf{r},t)^{\dagger}\hat{\sigma}^k\psi_j^{\mathrm{KS}} + \psi_j^{\mathrm{KS}}(\mathbf{r},t)^{\dagger}\hat{\sigma}^k\nabla\psi_j^{\mathrm{KS}}\big],
\end{equation}
that leads to
\begin{equation}
\mathcal{F}_{ik}(\mathbf{r},t) = \frac{1}{n}\sum_{j,r\in\mathrm{occ.}}\big[ \psi_r^{\mathrm{KS}\dagger}\hat{\sigma}^i\psi_r^{\mathrm{KS}}\nabla\psi_j^{\mathrm{KS}\dagger}\hat{\sigma}^k\psi_j^{\mathrm{KS}} + \psi_r^{\mathrm{KS}\dagger}\hat{\sigma}^i\psi_r^{\mathrm{KS}}\psi_j^{\mathrm{KS}\dagger}\hat{\sigma}^k\nabla\psi_j^{\mathrm{KS}}\big].
\end{equation}
The anti-symmetric part, $\mathcal{K}_{ik}(\mathbf{r},t)$, of the tensor $\mathcal{F}_{ik}(\mathbf{r},t)$, defined as $\mathcal{K}_{ik}=\mathcal{F}_{ik} - (i\leftrightarrow k)$ may be written as
\begin{equation}
\mathcal{K}_{ik}(\mathbf{r},t) = \frac{1}{n}\sum_{j,r\in\mathrm{occ.}}\sum_{\alpha,\beta,\alpha',\beta'}\big[ \psi_{r,\alpha}^{\mathrm{KS}*}\psi_{r,\beta}^{\mathrm{KS}}\sigma_{\alpha,\beta}^{[i, }\sigma_{\alpha',\beta'}^{k]}\nabla\psi_{j,\alpha'}^{\mathrm{KS}*}\psi_{j,\beta'}^{\mathrm{KS}} + \psi_{r,\alpha}^{\mathrm{KS}*}\psi_{r,\beta}^{\mathrm{KS}}\sigma_{\alpha,\beta}^{[i, }\sigma_{\alpha',\beta'}^{k]}\psi_{j,\alpha'}^{\mathrm{KS}*}\nabla\psi_{j,\beta'}^{\mathrm{KS}}\big],
\end{equation}
By making use of the following relation between Pauli matrices [see Ref.~\onlinecite{Taka55}]
\begin{equation}
\sigma_{\alpha,\beta}^{[i, }\sigma_{\alpha',\beta'}^{k]} = i\sum_s\epsilon_{iks}[ \sigma^s_{\alpha,\beta'}\delta_{\alpha',\beta} - \delta_{\alpha,\beta'}\sigma^s_{\alpha',\beta}],
\end{equation}
we obtain the following final expression for $\mathcal{K}_{ik}$ that can be splitted in two parts as follows
\begin{equation}\label{Eq:Kik}
\mathcal{K}_{ik}(\mathbf{r},t) = \sum_{j,r\in\mathrm{occ.}}\mathcal{K}_{ik}^{(j,r)}(\mathbf{r},t)\delta_{j,r} + \sum_{j\in\mathrm{occ.}}\sum_{r\neq j\in\mathrm{occ.}}\mathcal{K}_{ik}^{(j,r)}(\mathbf{r},t),
\end{equation}
where we have introduced the tensor
\begin{equation}
\mathcal{K}_{ik}^{(j,r)}(\mathbf{r},t) = \frac{i}{n}\sum_s\epsilon_{iks}\Big[ \psi_r^{\mathrm{KS}\dagger}\psi_j^{\mathrm{KS}}\big( \psi_j^{\mathrm{KS}\dagger}\hat{\sigma}^s\nabla\psi_r^{\mathrm{KS}} - \nabla\psi_j^{\mathrm{KS}\dagger}\hat{\sigma}^s\psi_r^{\mathrm{KS}}\big) + \psi_r^{\mathrm{KS}\dagger}\hat{\sigma}^s\psi_j^{\mathrm{KS}}\big( \nabla\psi_j^{\mathrm{KS}\dagger}\psi_r^{\mathrm{KS}} - \psi_j^{\mathrm{KS}\dagger}\nabla\psi_r^{\mathrm{KS}}\big) \Big].
\end{equation}
The procedure that we have followed up to now is formally exact. Then, in order to simplify the previous expression 
we substitute the Kohn-Sham ratio $\mathcal{F}_j=\frac{\psi_j^{\mathrm{KS}\dagger}\psi_j^{\mathrm{KS}}}{n(\mathbf{r},t)}$ 
with its average over the various occupied states $\mathcal{F}_j\simeq\bar{\mathcal{F}}=\frac{\langle\psi_j^{\mathrm{KS}\dagger}\psi_j^{\mathrm{KS}}\rangle_j}{n(\mathbf{r},t)}$. From the fact that, to a good degree of approximation, 
$\bar{\mathcal{F}}\simeq\frac{1}{N}$ with $N$ total number of particles in the system, we will
consider from now on $\bar{\mathcal{F}}$ to be spatially homogeneous and constant in time. 
Then Eq.~(\ref{Eq:Kik}) becomes
\begin{align}\label{Eq:Kik2}
\mathcal{K}_{ik}(\mathbf{r},t) & = i\bar{\mathcal{F}}\sum_s\epsilon_{iks}\sum_{j\in\mathrm{occ.}}\big[ \psi_j^{\mathrm{KS}\dagger}\hat{\sigma}^s\nabla\psi_j^{\mathrm{KS}} - \nabla\psi_j^{\mathrm{KS}\dagger}\hat{\sigma}^s\psi_j^{\mathrm{KS}}\big] + \frac{2m\bar{\mathcal{F}}}{\hbar}\sum_s\epsilon_{iks}\sum_{j\in\mathrm{occ.}}\psi_j^{\mathrm{KS}\dagger}\hat{\sigma}^s\psi_j^{\mathrm{KS}}\Big[\mathbf{v}_j(\mathbf{r},t) + \frac{e}{mc}\mathbf{A}(\mathbf{r},t)\Big] + \nonumber \\
& + \sum_{j\in\mathrm{occ.}}\sum_{r\neq j\in\mathrm{occ.}}\mathcal{K}_{ik}^{(j,r)}(\mathbf{r},t)\:.
\end{align}
In order to simplify the formalism we employ the notation $\mathcal{K}_{ik;l}(\mathbf{r},t) = n(s_i \partial_ls_k - s_k\partial_ls_i)$. Immediately from Eq.~(\ref{Eq:Kik2}) follows that
\begin{equation}
\frac{\hbar}{2\bar{\mathcal{F}}m}n\big(s_i \partial_ls_k - s_k\partial_ls_i\big) = -\sum_s\epsilon_{iks}\mathbf{J}_{\mathrm{KS}}^{sl}(\mathbf{r},t) + \sum_s\epsilon_{iks}\sum_{j\in\mathrm{occ.}}m_j^s(\mathbf{r},t)\Big[v_j^l(\mathbf{r},t) + \frac{e}{mc}A^l(\mathbf{r},t)\Big] + \frac{\hbar}{2\bar{\mathcal{F}}m}\sum_{\substack{j\in\mathrm{occ.}\\ r\neq j\in\mathrm{occ.}}}\mathcal{K}_{ik;l}^{(j,r)}(\mathbf{r},t)\:,
\end{equation}
where $\mathbf{m}_j$ and $\mathbf{v}_j$ define, respectively, the single Kohn-Sham state magnetization and velocity field. By 
employing the properties of the Levi-Civita tensor we have
\begin{equation}\label{Eq:3}
\frac{\hbar}{2\bar{\mathcal{F}}m}\big(n\mathbf{s}\times\partial_l\mathbf{s}\big)^n = -\mathbf{J}_{\mathrm{KS}}^{nl}(\mathbf{r},t) + \sum_{j\in\mathrm{occ.}}m_j^n(\mathbf{r},t)\Big[v_j^l(\mathbf{r},t) + \frac{e}{mc}A^l(\mathbf{r},t)\Big] + \mathcal{D}_{nl}(\mathbf{r},t)\:,
\end{equation}
where we have introduced the new tensor quantity
\begin{equation}\label{Eq:spindiss}
\mathcal{D}(\mathbf{r},t) = -\sum_{j\in\mathrm{occ.}}\sum_{r\neq j\in\mathrm{occ.}}\bigg[\mathcal{F}_{rj}\mathcal{J}^{(j,r)}(\mathbf{r},t) - \mathcal{F}_{jr}\mathbf{m}^{(r,j)}(\mathbf{r},t)\otimes\bigg(\mathbf{v}^{(j,r)}(\mathbf{r},t) + \frac{e}{mc}\mathbf{A}(\mathbf{r},t)\bigg)\bigg]\:.
\end{equation}
Here $\mathcal{F}_{rj}=\frac{\psi_r^{\mathrm{KS}\dagger}\psi_j^{\mathrm{KS}}}{n(\mathbf{r},t)}$ and the other many-particle 
objects are defined as
\begin{eqnarray}
\mathcal{J}^{(j,r)}(\mathbf{r},t) & = & -\frac{i\hbar}{2m}\big[ \psi_j^{\mathrm{KS}\dagger}\hat{\boldsymbol{\sigma}}\nabla\psi_r^{\mathrm{KS}} - \nabla\psi_j^{\mathrm{KS}\dagger}\hat{\boldsymbol{\sigma}}\psi_r^{\mathrm{KS}}\big]\:, \\
\mathbf{v}^{(j,r)}(\mathbf{r},t) & = & \frac{\hbar}{2mi}\frac{\psi_j^{\mathrm{KS}\dagger}\nabla\psi_r^{\mathrm{KS}} - \nabla\psi_j^{\mathrm{KS}\dagger}\psi_r^{\mathrm{KS}}}{\psi_j^{\mathrm{KS}\dagger}\psi_r^{\mathrm{KS}}} - \frac{e}{mc}\mathbf{A}(\mathbf{r},t)\:, \\
\mathbf{m}^{(j,r)}(\mathbf{r},t) & = & \psi_j^{\mathrm{KS}\dagger}\hat{\boldsymbol{\sigma}}\psi_r^{\mathrm{KS}}\:.
\end{eqnarray}
Finally, from Eq.~(\ref{Eq:3}) the divergence of the spin current tensor may be rewritten as
\begin{equation}\label{Eq:4}
-\nabla\cdot\mathbf{J}_{\mathrm{KS}}(\mathbf{r},t) = \frac{\hbar}{2\bar{\mathcal{F}}m}\nabla\cdot\big(n\mathbf{s}\times\nabla\mathbf{s}\big) - \sum_{j\in\mathrm{occ.}}\sum_l\partial_l\big[ \mathbf{m}_j(\mathbf{r},t)\cdot v_j^l(\mathbf{r},t)\big] - \frac{e}{mc}\sum_l\partial_l\big[ n\mathbf{s}A^l(\mathbf{r},t)\big] - \nabla\cdot\mathcal{D}(\mathbf{r},t)\:.
\end{equation}
Then, by substituting Eq.~(\ref{Eq:4}) into Eq.~(\ref{eq:SC}) we obtain
\begin{equation}
\frac{d}{dt}\mathbf{m}(\mathbf{r},t) = -\nabla\cdot\mathcal{D}(\mathbf{r},t) - \sum_{j\in\mathrm{occ.}}\sum_l\partial_l\big[\mathbf{m}_j(\mathbf{r},t)\cdot v_j^l(\mathbf{r},t)\big] + \frac{\hbar}{2\bar{\mathcal{F}}m}\nabla\cdot\big(n\mathbf{s}\times\nabla\mathbf{s}\big) + \mu_B n\mathbf{s}\times\mathbf{B}_{\mathrm{xc}}(\mathbf{r},t) + \mathbf{T}_{\mathrm{SO}}(\mathbf{r},t)\:.
\end{equation}
Finally, by decomposing the magnetization into its single-particle components, $\mathbf{m}_j$, we can define the 
magnetization material derivative as follows
\begin{equation}
\frac{D}{Dt}\mathbf{m}(\mathbf{r},t) = \frac{d}{dt}\sum_{j\in\mathrm{occ.}}\mathbf{m}_j(\mathbf{r},t) + \sum_{j\in\mathrm{occ.}}\big(\mathbf{v}_j\cdot\nabla\big)\mathbf{m}_j(\mathbf{r},t)\:,
\end{equation}
with the spin continuity equation that becomes
\begin{equation}
\frac{D}{Dt}\mathbf{m}(\mathbf{r},t) = -\nabla\cdot\mathcal{D}(\mathbf{r},t) - \sum_{j\in\mathrm{occ.}}\nabla\cdot\mathbf{v}_j(\mathbf{r},t)\mathbf{m}_j(\mathbf{r},t) + \frac{\hbar}{2\bar{\mathcal{F}}m}\nabla\cdot\big(n\mathbf{s}\times\nabla\mathbf{s}\big) + \mu_B n\mathbf{s}\times\mathbf{B}_{\mathrm{xc}}(\mathbf{r},t) + \mathbf{T}_{\mathrm{SO}}(\mathbf{r},t)\:,
\end{equation}
or
\begin{equation}
\frac{D}{Dt}\mathbf{m}(\mathbf{r},t) + \sum_{j\in\mathrm{occ.}}\nabla\cdot\mathbf{v}_j(\mathbf{r},t)\mathbf{m}_j(\mathbf{r},t) = -\nabla\cdot\mathcal{D}(\mathbf{r},t) + \mu_B\mathbf{m}(\mathbf{r},t)\times\mathbf{B}_{\mathrm{eff}}(\mathbf{r},t) + \mathbf{T}_{\mathrm{SO}}(\mathbf{r},t)\:,
\end{equation}
where we have introduced an effective magnetic field
\begin{equation}\label{Eq:Beff}
\mathbf{B}_{\mathrm{eff}}[n,\mathbf{s}](\mathbf{r},t) = \mathbf{B}_{\mathrm{xc}}[n,\mathbf{s}](\mathbf{r},t) + \frac{1}{\bar{\mathcal{F}}e}\bigg[ \frac{\nabla n\cdot\nabla\mathbf{s}}{n} + \nabla^2\mathbf{s}\bigg]\:.
\end{equation}
The continuity equation for the electron density instead follows immediately from Eq.~(\ref{eq:contn}) through the definition 
of velocity field
\begin{equation}
\frac{Dn}{Dt} = -n \nabla\cdot\mathbf{v}\:.
\end{equation}
It should be noted that the kinetic field, 
$\frac{1}{\bar{\mathcal{F}}e}\big[ \frac{\nabla n\cdot\nabla\mathbf{s}}{n} + \nabla^2\mathbf{s}\big]$, written in Eq.~(\ref{Eq:Beff}) 
is expressed in term of the density and spin density, that are observables of the many-body system. This means that the kinetic 
field obtained within the Kohn-Sham formalism is identical to its many-body counterpart. 
\end{widetext}

\end{document}